# MAGNETISM IN GRAPHENE SYSTEMS[*]

ERJUN KAN, ZHENYU LI, JINLONG YANG[*†]

*Hefei National Laboratory for Physical Sciences at Microscale, University of Science & Technology of China, Hefei, Anhui 230026, P. R. China*

*jlyang@ustc.edu.cn*

Graphene has attracted a great interest in material science due to its novel electronic structrues. Recently, magnetism discovered in graphene based systems opens the possibility of their spintronics application. This paper provides a comprehensive review on the magnetic behaviors and electronic structures of graphene systems, including 2-dimensional graphene, 1-dimensional graphene nanoribbons, and 0-dimensional graphene nanoclusters. Theoretical research suggests that such metal-free magnetism mainly comes from the localized states or edges states. By applying external electric field, or by chemical modification, we can turn the zigzag nanoribbon systems to half metal, thus obtain a perfect spin filter.

*Keywords*: Graphene; first-principles calculations; magnetism; half metal.

## 1. Introduction

Metal-free magnetism is the subject of an intense research, because of the small spin-orbit coupling and long spin scattering length. Due to the possibility of high Cure temperature, carbon-based nanostructures hold significant promise for future electronic devices.[1] Following the detection of room-Temperature weak ferromagnetism in polymerized C60,[1] several experimental groups have discovered magnetism in pure carbon materials.[2-4] Intrinsic carbon defects or adatoms are believed to be the origin of the unexpected magnetism.[5-7]

Recently, due to the progress of device setup techniques, a new kind of carbon materials named as graphene, has been fabricated.[8] Graphene is a monolayer of carbon atoms packed into a dense honeycomb crystal structure, which can be obtained by mechanical exfoliation from graphite. Many experimental studies have focused on the anomalous quantum hall effect (QHE).[9-13] Another important observation is that the electron mobility of graphene is about ten times higher than the mobility of commercial silicon wafers and electrons can travel huge distances (300 nm or more) without being scattered.[12-13] These excellent properties make graphene a potential substitute of silicon in electronics.

Because of the long spin-relaxation length and ballistic transport characteristics, graphene provides a great arena to develop the spin-polarized devices. Different with the *d* or *f* shell elements, carbon atoms themselves don't own magnetic moments. Therefore, the researches on grapheme based spintronics mainly pay attention to the substantial magnetism in graphene. In this review paper, we discuss three classes of graphene systems according to their dimensions: graphene, graphene nanoribbons (GNR), and graphene nanoclusters (GNC).

## 2. Graphene

Graphene, as a metal-free material, contains no magnetic atoms. Its honeycomb structure contains a bipartite lattice, formed by two interpenetrating triangular sublattics (A and B). The electronic structure of graphene can be described by a single-orbital nearest-neighbor hopping Hamiltonian.[14,15] This model correctly describes the graphene with linear bands around the Fermi energy. The magnetism in graphene comes from the local states caused by defects or molecular adsorption.

### 2.1. *Defects*

Defects in ideal graphene can be introduced by both vacancies and external doping. Many experimental works have reported the existence of magnetism in carbon materials by electrons or ions irradiation.[16-18] The common feature of these defects is that carbon atoms are removed from the graphene sheet, which gives quasilocalized states at the Fermi level.[19-20] First-principles calculations and tight-binding method have been applied for such systems.[21-24] Using density functional theory (DFT) with the generalized gradient approximation (GGA), Yazyev *et al.* have studied the magnetic behaviors of signal atom vacancies in graphene.[21] Calculated magnetic moments are equal to 1.12-1.53$\mu_B$ per vacancy depending on the defect concentration. As shown in Fig 1, vacancy induces an impurity state around the Fermi level, and break the $D_{3h}$ symmetry. The two sublattices have different spin populations. They also argued that only when these defects are produced in the same sublattice, the magnetic moments from the defective states can make a ferromagnetic (FM) coupling. Other groups have obtained similar conclusions.[22-24] Besides, Lehtinen *et al.* also found that additional hydrogen adsorption on the vacancy can destroy the magnetism.[22] On the other hand, Zhang *et.al.* concluded that the presence of nitrogen around a vacancy can produce larger macroscopic magnetic signals as compared to a standalone carbon vacancy.[23] The different electronic and magnetic behaviors of carbon vacancy under external elements adsorption provide a very valuable method to manipulate the magnetism of graphene.

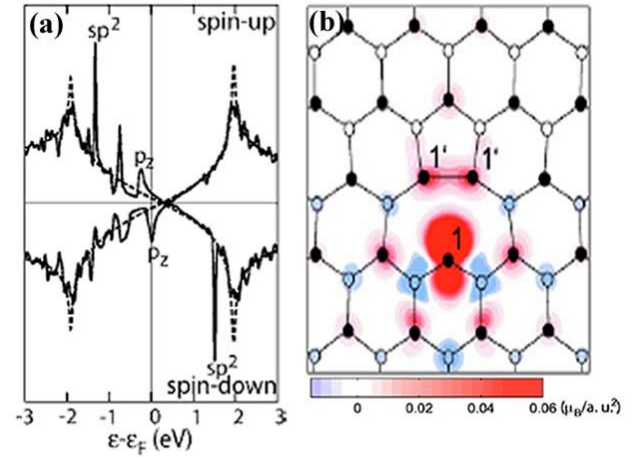

Fig. 1. (a) Density of states plots for the systems with the vacancy defects. The dashed line shows the density of states of the ideal graphene. Labels indicate the character of the defect states. (b) Spin-density projection (in $\mu_B$/a.u.$^2$) on the graphene plane around the vacancy defect. (From ref. 21)

As we know, besides the native defects (vacancies), graphene can also show magnetism by doping defects. Okada *et. al.* studied the electronic structures of hexagonally bonded honeycomb graphene by introducing B and N atoms.[25] According to their results, the π orbitals of the atoms aroud the border regions of graphite and BN are localized, and they are responsible for the magnetism. Theoretical calculations reveal that such magnetic atoms favor FM coupling. Their results provide an interesting direction to get metal-free FM materials.

### 2.2. *Adsorption*

Not only defects can produce magnetism, atom or molecule adsorptions also can lead to the occurrence of magnetic moments.

Yazyev *et al.*[21] and Boukhvalov *et.al.*[26] have studied the adsorption of hydrogen atoms on graphene. Their results confirmed that such adsorption will lead to magnetic moments on neighboring carbon atoms, and such spin-polarized states are mainly localized around the adsorptive hydrogen. Another feature is that the *sp$^2$* carbon atoms will become *sp$^3$* carbon, and make the graphene lose the $D_{3h}$ symmetry. Boukhvalov *et.al.*[26] also investigated the magnetic coupling under hydrogen pair. The calculated results show that only the hydrogen atoms distributing on the same sublattices can introduce FM coupling, while on the nearest carbon pairs, the dangling bonds of carbon atoms will be saturated, leading to non-magnetic system.

Other possible adsorptive atoms include carbon[27], nitrogen[28], and oxygen atoms[29]. The adsorptive carbon, nitrogen, and oxygen atoms prefer the bridge-like positions on graphene surface. Carbon and nitrogen atoms induce magnetic moments in the graphene, while oxygen atom cannot. The calculated diffusion barrier is 0.47 eV for carbon, and 1.1 eV for nitrogen.

## 3. Graphene nanoribbons

Graphene nanoribbons (GNRs) are grephene layer terminated in one direction with a specific widths. Their electronic structures have been broadly studied with tight-binding method and first-principles calculations.[30-34] The geometric structures of GNRs are shown in Fig. 2. Following the standard convention[30,33,34], armchair GNRs are classified by the number of dimmer lines ($N_a$) across the ribbon. Likewise, zigzag GNRs are classified by the number of the zigzag chains ($N_z$) across the ribbons.

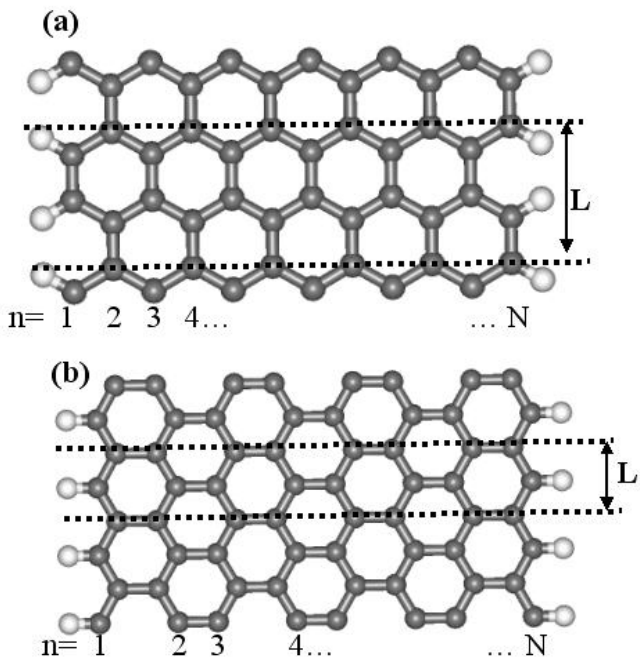

Fig. 2. Structures of (a) armchair graphene nanoribbons, and (b) zigzag grephene nanoribbons. Black balls are carbon atoms, and white balls are hydrogen atoms. The rectangle drawn with dashed lines denotes the unit cell.

Son *et.al.* have investigated the energy gaps of such GNRs with both modified tight-binding approximation and first-principles calculations.[33] Their results reveal that all such GNRs are semiconductors with an energy gap. For armchair GNRs, the energy gaps can be characterized by $\Delta \sim w_a^{-1}$, where $w_a$ means the width of a graphene nanoribbon. Similarly, the energy gaps can be fitted by $\Delta = 9.33/(w_z+15.0)$ for zigzag GNRs, $w_z$ means the width of a zigzag nanoribbon in angstrom. The most important thing is that edge states lead to the appearance of magnetic order in zigzag nanoribbons. Detailed theoretical research confirmed that such spin polarization has FM coupling on the same edges, while anti-ferromagnetic (AFM) coupling between the two challenge edges.[33, 35]

### 3.1. Armchira graphene naoribbons

Since all the previous calculations suggest that armchair GNRs are non-magnetic, it is challenge to find a reliable way to gain magnetism in such nanoribbons. Our previous calculations show modified edges with different chemical groups don't lead to magnetic structures.[36]

As a natural counterpart of armchair GNRs, carbon nanotubes (CNT) provide many valuable insights for the magnetic research. To take full advantage of the long spin-relaxation length and ballistic transport characteristics, many groups have investigated the combo of transitional metals and CNT.[37-40] Besides the substantial magnetism, their results also show that such structures may display half-metallicity.[37]

Metal atomic chains can adsorb on the graphene surface, and may provide the needed magnetism. Therefore, we performed first-principles calculations on the electronic structures of atomic titanium chain on armchair GNRs.[41] We consider three different Ti concentration in our models, as plotted in Fig. 3. The calculated results show all Ti atomic chains prefer the edges of the armchair GNRs, and favor FM coupling. Compared with CNT, armchair GNRs can bind the Ti chains more strongly. Another important feature is that all such hybridized structures are metallic.

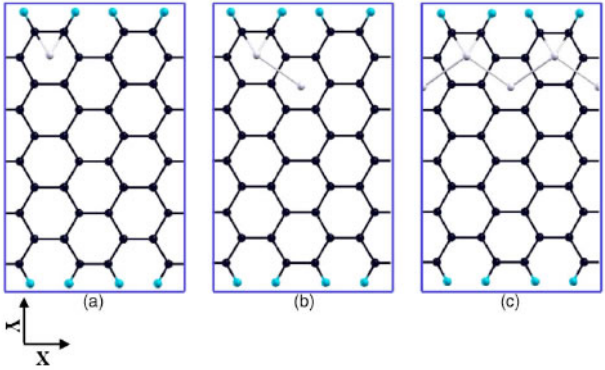

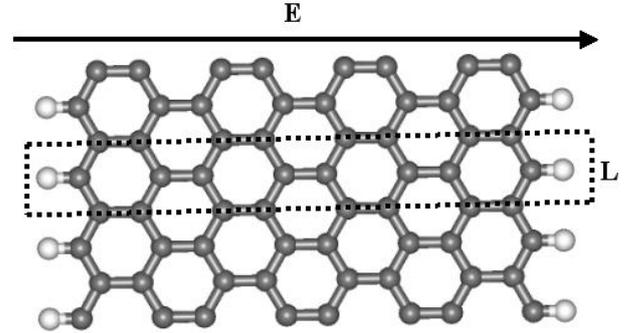

Fig.3. The structures of armchair GNRs adsorbed by Ti chains. Black balls are C atoms, sapphire balls are H atoms, and white balls represent Ti atoms. The rectangle drawn with a solid line denotes the cell which is double of the unit cell of pure armchair GNRs. (a), (b), and (c) represent the configurations in that the numbers of adsorbed transition-metal atoms per cell are one, two, and four. All structures are fully relaxed before performing band structure calculations. (From ref. 41)

Particularly, Our results also show that when the nanoribbons are less than 2.1 nm in width, the hybrid structures may present half-metallic behaviors. Thus, such hybrid structures between the metal atom chains and the graphene may provide an useful way to produce spintronics.

### 3.2. Zigzag graphene nanoribbons

It is now well-known that zigzag edge GNR is a semiconductor with two electronic edge states, which are FM ordered but AFM coupled to each other.[33-35] The magnetic edge states degenerate in energy. In the following, we show that there are several means to turn zigzag GRNs to half metal.

#### A) External electric fields

Because the magnetic states are the edge states, thus the external transverse electric fields are expected to have significant effect on such states. Son et. at.[42] have performed *ad initio* calculations based on density functional method within local spin density approximation. In their models, an external electric field is applied across the ribbons, as shown in Fig.4.

Fig.4. The structures of zigzag graphene nanoribbons: black balls are C atoms and white balls are H atoms. The rectangle drawn with dashed lines denotes the unit cell, and the arrow line represents the direction of external electric fields.

Their results reveal an interesting phenomenon: an applied electric fields removes the degenerate in energy of the two edges, and make the ribbons spin-selective. The break of degenerate is due to the additional potential caused by the electric fields. Fig. 5. shows the band structures of zigzag GRNs with a width of $N_z$ = 16 under the electric fields.

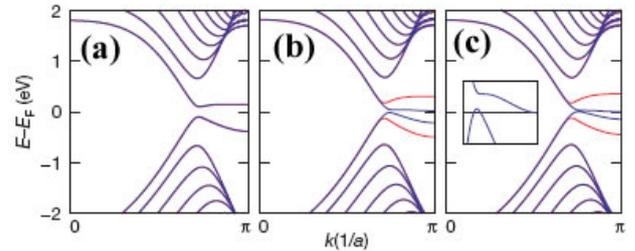

Fig. 5. The band structures of zigzag GNRs with $N_z$ = 16 under external electric fields. From left (a) to right (c), the spin-resolved band structures are response for electric fields with 0.0, 0.05, and 0.1 V/Å, respectively. The red and blue lines denote bands of α-spin and β-spin states. (From ref. 42)

With further increase of the electric field strength, the same spin states overlap each other, as shown in Fig.5(c). The above results mean that zigzag GNRs can be tuned into half-metallic material by electric fields. Importantly, such metal-free half metals may have high Cure temperature, thus have a great impact in the future electronic devices.

They also pointed out that the critical electric field to achieve half-metallicity in zigzag GNRs decreases as the width increases because the electrostatic potential difference between the two edges is proportional to the system size.[42] Their results provide an encouraging tactic to practical spintronics, because the electric fields can be controlled in application.

However, Rudberg *et.al.*[43] have argued that the half-metallicity should be removed by the non-local exchange interaction. In their research, they constructed a finite piece of zigzag GNRs, which contains 472 carbon atoms, and 74 hydrogen atoms, with $N_z$ = 8. Three different kind of functional have been adopted, i.e., LDA, GGA, and hybrid B3LYP. As shown in Fig. 6, their LDA and GGA results show that such zigzag GRNs can give half-metallic behaviors under external electric fields, similar with the results of Son's.[42] Surprisingly, within the framework of hybrid B3LYP, the half-metallicity is absent. The electronic structures are spin-selective semiconducting. Rudberg *et.al.*[43] have concluded that non-local exchange interaction is responsible for the lack of half-metallicity.

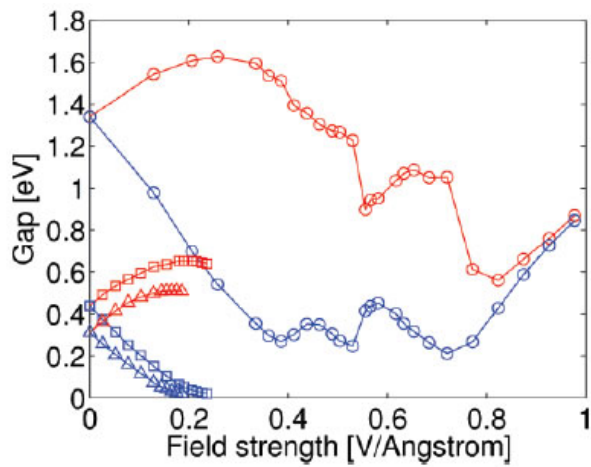

Fig.6. Computed band gaps plotted against external electric field, for a zigzag GNR with $N_z$ = 8. The band gaps were calculated using the LDA, BLYP, and B3LYP functionals and the 3-21G basis set. LDA, BLYP, and B3LYP results are shown as triangles, squares, and circles, respectively. (From ref. 43)

We have performed *ab initio* calculations within the framework of hybrid B3LYP functional to study the effect of the non-local exchange interaction on the infinite length zigzag GNRs,.[44] We adopted the zigzag GRNs with the same width ($N_z$) of Rudberg *et.al.*[43] Our calculated results are plotted in Fig.7. It is clear that our model can be turned into half metal under strong enough external electric fields.

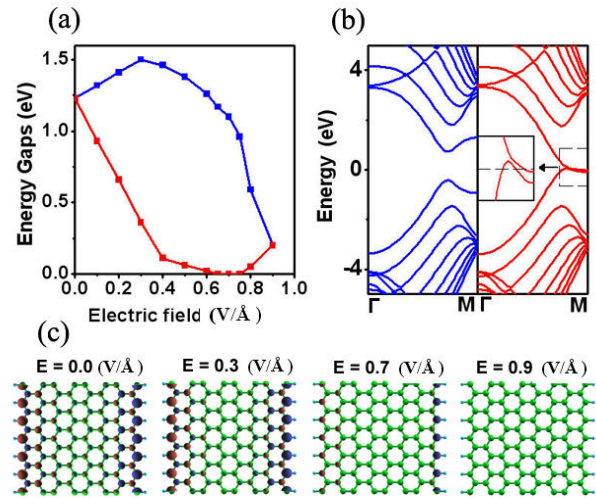

Fig. 7 (a) Spin-up (blue) and spin-down (red) zigzag GNR with $N_z$ = 8 band gaps against external electric fields. (b) Spin-up (blue) and spin-down (red) zigzag GNR with $N_z$ = 8 and structures with $E$=0.65 V/Å. (c) The spin densities of zigzag GNR with $N_z$ = 8 under different external electric fields: red for positive values and blue for negative. (From ref. 44)

Our results also reveal another feature: the half-metallic behaviors can only be realized in limited range of external electric fields. In order to get enough information about this property, we plotted the density of states (DOS) in Fig.8. (a) and (b). The DOS analysis shows that when the spin-polarized electrons of one edge have been fully transferred to the other one driven by the electric field, the magnetism is removed.

We also studied the relation between the widths of ribbons and the critical and range of electric field for achieving half-metallicity. As shown in Fig.8.(c) and (d), the critical field is inversely proportional to the width of ribbons, while the range is proportional to the width.

According to our results, non-local exchange interaction cannot remove the half-metallicity in infinite length zigzag GNRs, but indeed increases the critical electric field.

### B) Edge modification

Although the external electric field can tune the ribbons into half metals, it is a great challenge to apply strong field in large-scale applications. Since the half-metallicity is caused by the charge transfer, chemical decoration or modification may provide an alternative way to convert zigzag GNR to half metal.

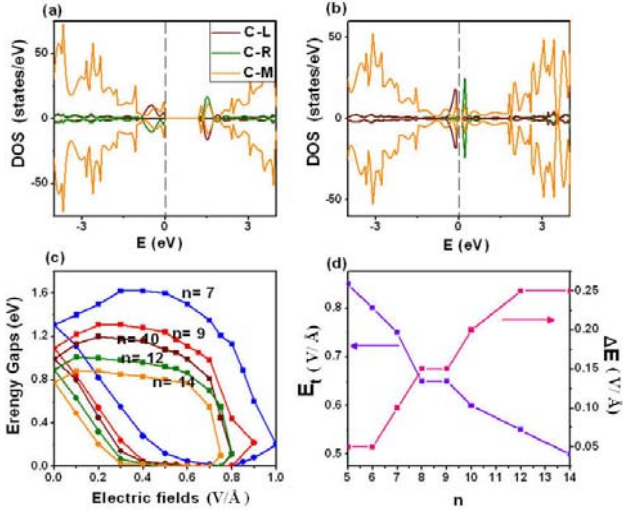

Fig. 8 PDOS of zigzag GNR with $N_z = 8$ under external electric field of (a) 0.0 V/Å and (b) 0.9 V/Å, positive for spin up and negative for spin down. (c) *zigzag* GNR band gaps against external electric fields for $N_z = 7, 9, 10, 12,$ and 14; the line with squares represents spin-up channel, and filled circles for spin-down one. (d) The critical electric fields ($E_t$) to achieve half-metallicity and the range of electric field strength (from $E_t$ to $E_t + \Delta E$) to keep half-metallicity for ribbons with $N_z = 5, 6, 7, 8, 9, 10, 12,$ and 14. (From ref. 44)

Gunlycke et.al. have studied the effect of different atoms or functional groups on electronic structures of zigzag GNRs.[45] They suggest that such edge-modification can alter the electronic structures. The carbon π orbitals at the edges were found to shift under effective potential induced by functional groups.

Another interesting result was reported by Hod *et.al*.[46] They studied the edge-oxidized zigzag graphene nano ribbons by hydroxyl, lactone, ketone, and ether groups. They find that these oxidized ribbons are more stable than hydrogen-terminated nanoribbons except in the case of the etheric groups. More importantly, the stable oxidized nanoribbons preserve the spin-polarized solutions with AFM coupling. Interestingly, they find that such edge oxidation can lower the onset electric field required to induce half-metallic behavior and extend the field range in which the systems remain half-metallic.[46]

As illustrated in Fig. 9, when the zigzag GNRs are modified with different groups at the two edges, the corresponding potential shifts are different. Thus, zigzag GNRs are expected to become spin-selective materials. Once the potential difference is large enough, conduction band (CB) and valence band (VB) in one spin channel (spin-down channel in Fig. 9) will overlap in energy, and the zigzag GNR is expected to become half metal.

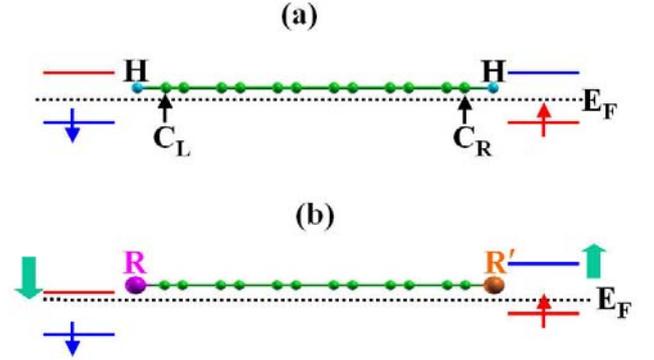

Fig. 9 (a) Structures of the H-saturated zigzag GNR and schematic energy diagram for the edge states. Green and sapphire balls denote carbon and hydrogen atoms, respectively. (b) Schematic edge-state energy diagram of zigzag GNR with different chemical modifications at the two edges. R (magenta) and R' (orange) represent different functional groups. Red energy levels for the spin-up channel, and blue for the spin-down channel. (From ref. 47)

To confirm our hypothesis, we have performed first-principles calculations.[47] In our calculations, we considered different functional groups. We denoted it as A-B when the zigzag GNRs are modified with A groups at one edge, and B groups on the other edge. The considered pairs include $NO_2$-$NH_2$, $NO_2$-H, and $NO_2$-$CH_3$. The calculated electronic structures show that $NO_2$-$NH_2$ pair tunes the zigzag GNRs into spin-selective metals, while $NO_2$-H pair converts the ribbons into spin-selective semiconductors. The most important result is that $NO_2$-$CH_3$ pair can make the ribbons become half metal. As plotted in Fig.10, we find that all the ribbons with different width included in our calculations can show half-metallicity through the modification of the $NO_2$-$CH_3$ pair.

As a critical factor in practice, the relative stability of the edge-modified zigzag GNRs is very important. Because these structures have different chemical compositions, the binding energy per atom does not provide a suitable measurement for the comparison of their relative stability. Therefore, we define a Gibbs free energy of formation δG for edge-modified zigzag GNRs as: δG= $E_c - n_H\mu_H - n_O\mu_O - n_N\mu_N - n_C\mu_C$, where $E_c$ is the cohesive energy per atom of chemically functionalized zigzag GNRs, $n_i$ is the molar fraction of atom i ( i = C, O, H, N) in the ribbons, satisfying the relation $n_H+n_O+n_N+n_C=1$. The binding energy per atom of $H_2$, $N_2$, and $O_2$ molecules are chosen as $\mu_H$, $\mu_O$, and $\mu_N$, respectively. And $\mu_C$ is the cohesive energy per atom of the infinite graphene. Our calculations reveal zigzag GNR modified by one $NO_2$-$CH_3$ pair per zigzag GNR unit (ZGNR-full) is less stable, and has a larger δG than the H-saturated one. This is easy to understand since the nearest-neighboring $NO_2$ and $CH_3$ group repulsion is too large at short distances. To obtain more stable products, we construct a new structure, where there is only one $NO_2$ and $CH_3$ pair every two unit cells (ZGNR-half). The calculated Gibbs free energy of formation is greatly reduced, and ZGNR-half is even more stable than H-saturated ZGNR. The energy gaps of ZGNR-half are also presented in Fig.10. We clearly find that such structures are also half metals with $N_z$ larger than 12. Thus, edge modifications should provide a possible way to produce zigzag GNR based spintronic devices.

expect that such hybrid structures can lead to different potential at the two edges, and form half metal. The spin-polarized DFT calculations confirm our guess, and show that half-metallicity can be presented with enough width of the ribbons in all three hybrid structures in our research. We find this unexpected half-metallicity in the hybrid nanostructures stems from a competition between the charge and spin polarizations, as well as from the π orbital hybridization between C and BN. The molecular-dynamics simulations show that such materials are stable under the room temperature.

Nakamura et.al. reported their research on the $(BN)_1C_{2n}$ and $(NB)_1C_{2n}$ ribbons.[51] The geometric structures are shown in Fig.11. Their results show that such structures favor ferrimagnetic states, with the spin density mainly distributes on the carbon edge. These structures present the magnetic moments, may give direct for production of macroscopical magnetic moments. Another important result is the calculated band structures of the $(BN)_1C_{2n}$ and $(NB)_1C_{2n}$ ribbons are metallic.

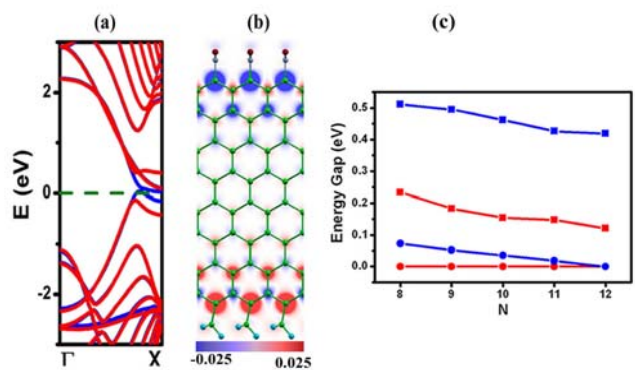

Fig.10. (a) Band structure and (b) spin density of zigzag GNRs modified by $NO_2$-$CH_3$ pair with $N_z$ = 8. Red represents the spin-up channel, and blue for the spin-down channel. (c) The band gaps of ZGNR-full (red) and ZGNR-half (blue) with spin-up (squre) and spin-down (circle) channels. (From ref. 47)

### C) Chemical doping

As a counterpart of edge modification, chemical doping method has been investigated.[48, 51] Native doping elements are boron and nitrogen. Keeping in mind that BN zigzag nanoribbons have the similar structures with C,[49, 50] we designed a new hybrid structure named as C/BN nanoribbons in our recent research.[48] Our structures constitute $C_iBN$, ( i= 1,2,3). Similar with the edge-modified zigzag GNRs, we

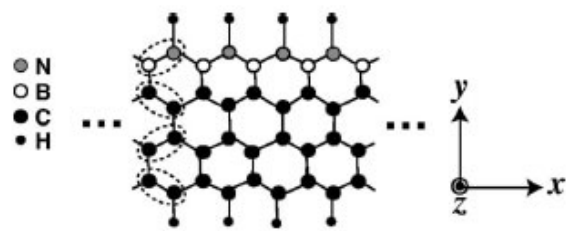

Fig.11. The atomic configuration of the $(NB)_1(C2)_3$ ribbon. Black, shaded, and white circles and small black dots denote C, N, B, and H atoms, respectively. (From ref. 51)

The authors also discussed why the spin density of the ferrimagnetic ground state appears at the C edge, irrespective of the ribbon structure, $(BN)_1C_{2n}$ or $(NB)_1C_{2n}$? In their schematic energy diagram (shown in Fig.12), they argued that the border state between B and C atoms is the bonding C state (*b*-C in Fig. 12). Because the C edge states are higher in energy than the bonding C state, thus the bonding C state is fully occupied for both the majority and minority spin channels. While only the majority spin channel of the C edge state is occupied. Therefore, the spin density only appears at the C edge. On the other hand, for $(BN)_1C_{2n}$ ribbons, the interaction between N and C atoms forms the antibonding C state (*a*-C in Fig. 12). Since the antibonding C state is higher in energy than that of C edge state, this antibonding orbital should not be occupied. While for the C edge states, electrons in majority spin channel is much more than that in minority spin one. Thus the total spin density appears at the C edge.

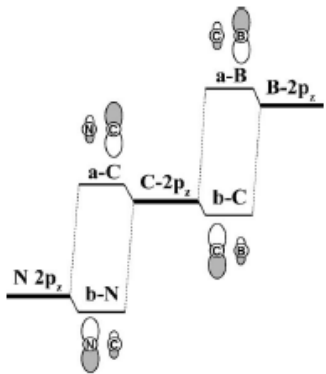

Fig.12. Schematic of the energy diagram for the formation of the border state. (From ref. 51.)

### 4. Graphene nano clusters (GNC)

Graphene nano clusters (GNC) are finite fragments cut from graphene, which lose translational symmetry. According to their special shapes, they have been named as nanoislands [52] and nanoflakes.[53] Since such nano structures are difficult to be realized, theoretical researches can help us to gain enough insights into the mechanics of magnetic stability. Different with the zigzag GNRs, the magnetism is from the nonbonding states (NBS) caused by the zigzag edge atoms.[53] Total magnetic moments can be described by a formula: $2S = N_A - N_B$, where $N_A$, $N_B$ are atoms belonging to different sublattices.[52, 53] As a results, when the nanoislands or nanoflakes have the same number of sublattices atoms, such structures have no net magnetic moments.[52] But we must notice that no net magnetic moments does not mean such structures are nomagnetic. Just as the results of Fernández-Rossier *et.al.*[52], each egde atoms still have a net magnetic moment, but the AFM coupling between the nearest neighbour edges of the nanoislands will remove the net magnetic moment of the system.

### 5. Summary

This paper provides a review on the magnetism of graphene systems from a theoretical perspective. The different magnetic origins in graphene, GNRs, and GNC have been described. We also show that through different methods, such as applying external electric field, edge modification, and chemical doping, the magnetic behaviors and the electronic structures of zigzag GRNs can be remarkably tuned. The research of artifical GNC has also provided an avenue to nanoscale spintronics.

One of the present obstacles is how to realize controllable magnetism and high Cure temperature in such graphene systems. Because of the great potential of graphene in future electronic devices, such problems need further studied by experimental and theoretical works.

### Acknowledgments

This work is partially supported by the National Natural Science Foundation of China (50721091, 20533030, 50731160010), by National Key Basic Research Program under Grant No.2006CB922004, by the USTC-HP HPC project, and by the SCCAS and Shanghai Supercomputer Center.

**References**

1. T. L. Makarova, B. Sundqvist, R. Höhne, P. Esquinazi, Y. Kopelevich, P. Scharff, V. A. Davydov, L. S. Kashevarova, and A. V. Rakhmanina, *Nature (London)* 413, 716 (2001).
2. D. P. Esquinazi, A. Setzer, R. Höhne, C. Semmelhack, Y. Kopelevich, D. Spemann, T. Butz, B. Kohlstrunk, and M. Löche, *Phys. Rev. B* 66, 024429 (2002).
3. R. A. Wood, M. H. Lewis, M. R. Lees, S. M. Bennington, M. G. Cain, and N. Kitamura, *J. Phys. Condens. Matter* 14, L385 (2002).
4. J. M. D. Coey, M. Venkatesan, C. B. Fitzgerald, A. P. Douvalis, and I. S. Sanders, *Nature (London)* 420, 156 (2002).
5. P. O. Lehtinen, A. S. Foster, A. Ayuela, A. Krasheninnikov, K. Nordlund, and R. M. Nieminen, *Phys. Rev. Lett* 91, 017202 (2004).
6. P. O. Lehtinen, A. S. Foster, A. Ayuela, T. T. Vehvilinen, and R. M. Nieminen, *Phys. Rev. B* 69, 155422 (2004).
7. Yuchen Ma, P. O. Lehtinen, A. S. Foster, and R. M. Nieminen, *New J. Phys* 6, 68 (2004).
8. K. S. Novoselov, A. K. Geim, S. V. Morozov, D. Jiang, Y. Zhang, S. V. Dubonos, I. V. Grigorieva, and A. A. Firsov, *Science* 306, 666 (2004).
9. T. Ohta, A. Bostwick, T. Seyller, K. Horn, E. Rotenberg, *Science* 313, 951 (2006).
10. K. S. Novoselov, Z. Jiang, Y. Zhang, S. V. Morozov, H. L. Stormer, U. Zeitler, J. C. Maan, G. S. Boebinger, P. Kim, A. K. Geim, *Science* 315, 1379 (2007).
11. A. K. Geim, and K. S. Novoselov, *Nature Materials* 6, 183 (2007).
12. Y. Zhang, Y. Tan, H. L. Stormer, and P. Kim, *Nature* 438, 201 (2005).
13. K. S. Novoselov, A. K. Geim, S. V. Morozov, D. Jiang, M. I. Katsnelson, I. V. Grigorieva, S. V. Dubonos, and A. A. Firsov, *Nature* 438, 197 (2005).
14. R. Saito, M. S. Dresselhaus, and G. Dresselhaus, *Physical Properties of Carbon Nanotubes* (Imperial College Press, London, 1998).
15. P. K. Wallace, *Phys. Rev.* 71, 622 (1947).
16. P. Esquinazi, D. Spemann, R. Höhne, A. Setzer, K.-H. Han, and T. Butz, *Phys. Rev. Lett.* 91, 227201 (2003).
17. K. Urita, K. Suenaga, T. Sugai, H. Shinohara, and S. Iijima, *Phys. Rev. Lett.* 94, 155502 (2005).
18. A. Hashimoto, K. Suenaga, A. Gloter, K. Urita, and S. Iijima, *Nature (London)* 430, 870 (2004).
19. V. M. Pereira, F. Guinea, J. M. B. Lopes dos Santos, N. M. R. Peres, and A. H. Castro Neto, *Phys. Rev. Lett.* 96, 036801 (2006).
20. P. Ruffieux, M. Melle-Franco, O. Gröning, M. Bielmann, F. Zerbetto, and P. Gröning, *Phys. Rev. B* 71, 153403 (2005).
21. O. V. Yazyev, L. Helm, *Phys. Rev. B* 75, 125408 (2007).
22. P. O. Lehtinen, A. S. Foster, Y. Ma, A. V. Krasheninnikov, and R. M. Nieminen, *Phys. Rev. Lett.* 93, 187202 (2004).
23. Y. Zhang, S. Talapatra, S. Kar, R. Vajtai, S. K. Nayak, and P. M. Ajayan, *Phys. Rev. Lett.* 99, 107201 (2007).
24. H. Amara, S. Latil, V. Meunier, Ph. Lambin, and J.-C. Charlier, *Phys. Rev. B* 76, 115423 (2007).
25. S. Okada and A. Oshiyama, *Phys. Rev. Lett.* 87, 146803 (2001).
26. D. W. Boukhvalov, M. I. Katsnelson, A. I. Lichtenstein, *Phys. Rev. B* 77, 035427 (2008).
27. P. O. Lehtinen, A. S. Foster, A. Ayuela, A. Krasheninnikov, K. Nordlund, and R. M. Nieminen, *Phys. Rev. Lett.* 91, 017202 (2003).
28. Yuchen Ma, A. S. Foster, A. V. Krasheninnikov, and R. M. Nieminen, *Phys. Rev. B* 72, 205416 (2005).
29. D. C. Sorescu, K. D. Jordan, and P. Avouris, *J. Phys. Chem. B* 105, 11227 (2001).
30. M. Fujita, K. Wakabayashi, K. Nakada, and K. Kusakabe, *J. Phys. Soc. Jpn.* 65, 1920 (1996).
31. K. Nakada, M. Fujita, G. Dresselhaus, and M. S. Dresselhaus, *Phys. Rev. B* 54, 17 954 (1996).
32. M. Ezawa, *Phys. Rev. B* 73, 045432 (2006).
33. Y.-W. Son, M. L. Cohen, and S. G. Louie, *Phys. Rev. Lett.* 97, 216803 (2006).
34. L. Yang, C. H. Park, Y. W. Son, M. L. Cohen, and S. G. Louie, *Phys. Rev. Lett.* 99, 186801 (2007).
35. L. Pisani, J. A. Chan, B. Montanari, and N. M. Harrison, *Phys. Rev. B* 75, 064418 (2007).
36. Hao Ren, Qunxiang Li, Haibin Su, Q. W. Shi, Jie Chen, and Jinlong Yang, *cond-mat: 0711.1700v1*.
37. C. Yang, J. Zhao, and J. P. Lu, *Nano Lett.* 4, 561 (2004).
38. C. K. Yang, J. Zhao, and J. P. Lu, *Phys. Rev. Lett.* 90, 257203 (2003).
39. Viktoria V. Ivanovskaya, Christof Köhler, and Gotthard Seifert, *Phys. Rev. B* 75 075410 (2007).
40. Mariana Weissmann, Griselda García, Miguel Kiwi, Ricardo Ramírez, and Chu-Chun Fu, *Phys. Rev. B* 73 125435 (2006).
41. Er-Jun Kan, H. J. Xiang, Jinlong Yang, J. G. Hou *J. Chem. Phys.* 127 164706 (2007).
42. Y.-W. Son, M. L. Cohen, and S. G. Louie, *Nature (London)* 444, 347 (2006).
43. E. Rudberg, P. Salek, and Y. Luo, *Nano Lett.* 7, 2211 (2007).
44. Er-Jun Kan, Zhenyu Li, Jinlong Yang, J. G. Hou, *Appl. Phys. Lett.* 91, 243116 (2007).
45. D. Gunlycke, J. Li, J. W. Mintmire, and C. T. White, *Appl. Phys. Lett.* 91, 112108 (2007).
46. O. Hod, V. Barone, J. E. Peralta, and G. E. Scuseria, *Nano Lett.* 7, 2295 (2007).
47. Er-Jun Kan, Zhenyu Li, Jinlong Yang, J. G. Hou, *J. Am. Chem. Soc.* 130, 4224 (2008).
48. Er-Jun Kan, Xiaojun Wu, Zhenyu Li, X. C. Zeng, Jinlong Yang, J. G. Hou, *cond-matt:0803.2073*.
49. Z. Zhang, and W. Guo, *Phys. Rev. B* 77, 075403 (2008).
50. Veronica Barone, Juan E. Peralta, *cond-matt:0803.2024*.

51. J. Nakamura, T. Nitta, and A. Natori, *Phys. Rev. B* **72**, 205429 (2005).
52. J. Fernández-Rossier and J. J. Palacios, *Phys. Rev. Lett.* **99**, 177204 (2007).
53. W. L. Wang, S. Meng, and E. Kaxiras, *Nano Lett.* **8**, 241 (2008).